\documentclass[aps,prl,twocolumn]{revtex4}
\usepackage{amsmath}
\usepackage{graphicx}

\newcommand{\ket}[1]{|#1\rangle}

\newcommand{\eq}{\begin{equation}}
\newcommand{\fine}{\end{equation}}

\begin{document}

\title{Entanglement localization after a coupling to incoherent noisy system}
\author{Fabio Sciarrino$^{1}$, Eleonora Nagali$^{1}$, Francesco De Martini$^{1,2}$%
, Miroslav Gavenda$^{3}$, and Radim Filip$^{3}$ \\
$^{1}$Dipartimento di Fisica dell'Universit\'{a} ''La Sapienza'' and
Consorzio Nazionale Interuniversitario per le Scienze Fisiche della Materia,
Roma 00185, Italy\\
$^{2}$Accademia Nazionale dei Lincei, Italy\\
$^{3}$Department of Optics, Palack$\acute{y}$ University, 17. Listopadu 50,
Olomuc 77200, Czech Republic}

\begin{abstract}
We report the experimental realization of entanglement localization which restores polarization entanglement completely redirected after a linear coupling with incoherent and noisy surrounding photon. The method, based only on measurements of the surrounding photon after the coupling and on post-selection, can localize the entanglement back to original systems for any linear coupling. 
\end{abstract}

\maketitle

Quantum entanglement is the cornerstone of many
protocols, both of quantum communication and information processing \cite{Niel}. In the real world, it is a fragile resource easily spoiled by the coupling to another surrounding system. 
In the last few years various schemes have been implemented in order to overcome this problem: distillation, purification and concentration
protocols \cite{Pan03,Kwia01,Pete05}. All these techniques deal with an entanglement
recovery after a non-completely destructive interaction with noise in
the communication channel. However a wider control on the entanglement resource and its dynamics is desiderable also when the interaction with noise is strong enough to completely destroy the quantum correlation of the signal. In order to achieve this aim we shall exploit the concept of entanglement localization (EL), a physical resource connected to the correlation function \cite{Vers04,Popp05}. Within a many-particles system, the localizable entanglement is defined as the maximal amount of entanglement that can be localized in a sub-system (for example, two particles), by making local measurements on the rest of the system. In the last few years, the investigation of EL features has attracted much attention \cite{Amic08}, in particular entanglement localization has been mainly discussed for pure states and for some theoretical examples from a large variety of mixed states \cite{Vers04,Popp05}. 
In the framework of entanglement distribution, the EL can be considered as a novel method in the extreme case where the entanglement is completely redirected through an interaction to the surrounding systems. In the present paper we apply the concept of entanglement localization in the case of coupling to a surrounding system which destroys entanglement by transferring them to the surrounding system. We theoretically investigate and experimentally prove the entanglement localization after a basic linear coupling of a photon from a maximally-entangled polarization pair with a small surrounding system represented by a single photon, which is completely depolarized and fully incoherent. 

\begin{figure}[t]
\centering
\includegraphics[width=0.35\textwidth]{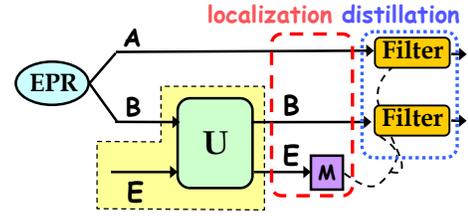}
\caption{ Schematic representation of the entanglement
localization procedure. The yellow dashed box indicates the inaccessible
surrounding system.}
\end{figure}


As a conceptual scheme, we consider the system presented in Fig.1. Our focus is on the simple physical example of interaction $U$ between two entangled qubits $A$ and $B$ (the signal), generated by an EPR source, and a small surrounding system represented by a single photon $E$ \cite{Niel,Zur}. Since there is no control on $E$ before its coupling to the system $B$, its quantum state is assumed to be completely mixed and to be incoherent. Incoherent means that $E$ is principally distinguishable from the signal photon. These two properties represents our novel focus in the entanglement localization.  After the interaction with the surrounding system, the initial entanglement between qubits $A$ and $B$ is lost. Here we will discuss how entanglement can be localized back to the communicating parties $A$ and $B$ by carrying out a proper measurement on the outgoing system $E$ and by a proper feed-forward quantum correction. The entanglement can be recovered through localization for any linear couplings. A maximum of the localized entanglement can be achieved by carrying out a probabilistic filtration. Moreover, we show that an induced partial indistinguishability between the entangled system $B$ and surrounding system $E$ can be exploited to improve the amount of localized entanglement. The analysis has been carried out for different strength of the coupling $T$ and by quantifying the entanglement by the councurrence $C$ \cite{Woot98}.

\begin{figure}[t]
\centering
\includegraphics[width=0.40\textwidth]{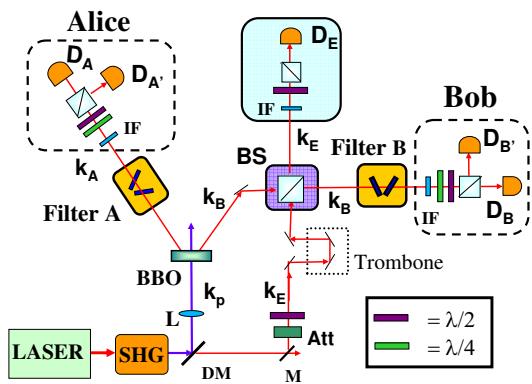}
\caption{ Experimental setup \protect\cite{Bouw97}. The main source of the experiment is a Ti:Sa
mode-locked laser with wavelength $\protect\lambda =795nm$.  A
small portion of the laser beam generates the single noise photon over the
mode $k_{E}$ using an attenuator ($Att$), which gives rise to a mean photon number equal to $\bar{n}=0.1$. Hence single photon contribution is achieved a posteriori. The main part of the laser beam passes through second harmonic generation (SHG)  where a bismuth borate
(BiBO) crystal generates a UV laser beam having
wave-vector  $k_{p}$ and wl $\lambda _{p}=397.5nm$ with power
equal to $800mW$. The transformation used to map the
state $\left| H\right\rangle _{E}$ into $\protect\rho _{E}=\frac{I_{E}}{2}$
is achieved through a stochastically rotated $\protect\lambda /2$ waveplate \cite{Scia04}.
The trombone is used to randomly shift the surrounding single photon $E$ out of the coherence time of the photon B.
The UV laser beam pumps a $1.5mm$ thick non-linear
crystal of $\protect\beta $-barium borate (BBO) cut for type II
phase-matching  which
generates polarization entangled pairs with equal wavelength $\protect\lambda =795nm$ \protect\cite{Kwia95}. The dashed
boxes indicate the polarization analysis setup adopted by Alice and Bob. The
photons are coupled to a single mode fiber and detected by single photon
counting modules $D_{i}$. On output modes $k_{B}$ and $%
k_{E}$ the photons are filtered adopting filters (IF) with $\Delta
\protect\lambda =3nm$ centered at $795nm$, while on mode $k_{A}$ $\Delta \protect\lambda =4.5nm$ for the distinguishable photons.}
\end{figure}

\textbf{Distinguishable surrounding photon:}
As first scenario we consider the case in which the photons $E$ and $B$ are in principle \textit{distinguishable} (incoherent). In this case, the two photons injected
into the beam-splitter ($BS$) have a random mutual delay $\Delta t$ such that $\Delta t\gg\tau _{coh}=300fs$, where $\tau_{coh}$ is the coherent time of the entangled pair. The resolution time of the detector satisfies $t_{\det }\gg \Delta t,$ hence it is not technologically possible to individuate whether the detected photon is $E$ or $B$.
The overall dynamic of the protocol can be divided in three different sequences. A
schematic drawing of our experimental layout is shown in Fig.(2).

\textbf{I) Coupling with surrounding system.} The localization protocol starts from a
 maximally entangled state $|\Psi ^{-}\rangle _{AB}=\frac{(|H\rangle
_{A}|V\rangle _{B}-i|V\rangle _{A}|H\rangle _{B})}{\sqrt{2}}$. During its propagation, the
photon $B$ is coupled with the surrounding photon $E$, described by the density matrix {\bf ${\rho }_{E}=\frac{{I}}{2}
$} corresponding to an unpolarized photon. It interacts with the photon $B$ by a linear unitary coupling: 
a beam splitter BS with transmittivity $T$ and reflectivity $R=1-T$ \cite{Gen}. After the interaction on the beam splitter, three possible situations can be observed:
both photons go to the measurement box or to Bob's detection, or only a single
photon is separately presented in both the detectors. The first case
corresponds simply to attenuation, while the second case can be in principle
distinguished by counting the number of photons in the signal. Thus only the
last case, which corresponds to a mixing of the photons $B$ and $E$, deserves interest. The output state reads:
\begin{equation}
\rho_{out}=q |\Psi ^{-}\rangle _{AB} \langle\Psi ^{-}| \otimes \frac{I_E}{2}+(1-q)|\Psi ^{-}\rangle _{AE} \langle\Psi ^{-}| \otimes \frac{I_B}{2}
\end{equation}
with $q=T^2/(T^2+R^2)$ connected to the strength of the coupling. The state $\rho_{AB}^{I}$ of the photon $A$ and $B$ is a Werner one and is found to be separable depending for $T<\sqrt{2}-1$: Fig.3 dotted line.

\begin{figure}[t]
\centering
\includegraphics[width=0.26\textwidth]{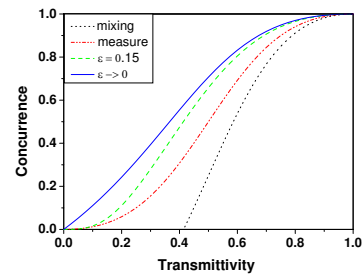}
\caption{Concurrence between $A$ and $B$ versus $T$ after the coupling with an incoherent unpolarized (distinguishable) photon $E$: before measurement on $E$ (dotted
line), after measurement on $E$ (dashed-dotted line), after measurement on $E$ and
filtration with $\protect\varepsilon =0.15$ on $A,B$ (dashed line), after measurement on $E$ and
filtration with $\protect\varepsilon \rightarrow 0$ on $A,B$ (continuous line). }
\end{figure}

\textbf{II) Entanglement localization.}
The measurement on the photon $E$ has been used to localize the entanglement between Alice and Bob. As no frequency selection has been performed on the laser mode $k_E$ before the coupling with signal, a
multi-mode single photon state interacts with the entangled photon $B$. We have measured
only a tiny part of frequency spectrum of this multi-mode light selected by a single mode fiber and an interference filter (IF) in front of detector $D_{E}$. In this way, we relax from the condition to measure all the interacting photons. The photon propagating on selected mode $k_{E}$ is then measured
after a polarization analysis (Fig.2). Hence the output state $\rho_{out}$ is projected onto the 
$|H\rangle _{E}$ \cite{Feed}:
\begin{equation}
\rho_{AB}^{II}=q|\Psi ^{-}\rangle _{AB} \langle\Psi ^{-}| + (1-q) |V\rangle _{A} \langle V| \otimes \frac{I_B}{2}
\end{equation}
This state is entangled for any value of $T$ as shown by Fig.3. The entanglement localization effect can be explained as follows: we exploit the correlation between the output photons $A$ and $E$ in Eq.2 (note that the photons $B$ and $E$ are completely uncorrelated since they do not overlap on the beamsplitter). The measurement on photon $E$ allows to perform a partial projection on $A$: by measuring the polarization of the photon $E$ the depolarized noise acting on the photon $A$ is transformed into a polarized noise (see transition from I to II), which is less destructive for entanglement. Furthermore, as will be shown below, the amount of entanglement can be probabilistically enhanced through a local procedure.

\begin{table}[t!]
\begin{center}
{\footnotesize
\begin{tabular}{||c||c|c||c|c||}
\hline\hline
$\rho_{AB}$ & $A_{A}$ & $A_{B}$ & $Concurrence$ & $P$ \\ \hline\hline
$I$ &
$$
&
$$
&  $Max(0,\frac{%
2T^2-R^2}{2(R^2+T^2)})$ & $R^2+T^2$ \\ \hline
$II$ &
$$
&
$$
&  $\frac{T^2}{T^2+R^2}$ & $%
\frac{1}{2}(R^2+T^2)$  \\ \hline
$III$ & $\frac{\varepsilon T^2}{T^2+R^2}$ & $\varepsilon $ &  $\frac{1}{1 + \varepsilon \frac{R^2}{2 T^2}}%
\frac{T}{\sqrt{T^2+R^2}}$ & $\frac{1}{2} \varepsilon T^{2} + \frac{1}{4} {%
\varepsilon}^2 R^2$ \\ \hline\hline
\end{tabular}
}
\end{center}
\caption{Entanglement localization for the linear coupling to distinguishable depolarized surrounding photon: concurrence C, probability of success P, and intensity $A$ of
filtering on modes $A$ and $B$, for each step of the
localization procedure. }

\end{table}


\textbf{III) Entanglement filtration.}
To enhance the localized entanglement after the measurement on the photon $E$, the probabilistic single-copy quantum filtration  introduced in \cite{Kwia01} can be used. 
To this scope, two filters $F_A$ and $F_B$
based on two sets of glass close to their Brewster's angle, were placed on Bob's and Alice's mode in order to
attenuate one polarization ($V$) in comparison to its orthogonal ($H$). We indicate the
attenuation over the mode $k_{i}$ for the $V$ polarization as $A_{i}$: $\ket{V}_{i}\rightarrow \sqrt{A_{i}(\varepsilon)}\ket{V}_{i}$  where $\varepsilon<$ is a parameter such that $0<\varepsilon<1$(see Table I).  By tuning the incidence angle of the beam on mode $k_{i}$, different values of
 $\varepsilon$ can be achieved . The following state is achieved

\begin{equation}
\begin{aligned}
\scriptstyle
\rho_{AB}^{III}=&\frac{1}{P_{III}}(\frac{\varepsilon T^3}{\sqrt{T^2+R^2}} |\Psi ^{-}\rangle _{AB} \langle\Psi ^{-}| + \varepsilon^2 R^2 |VV\rangle _{AB} \langle VV|+ \\
&\varepsilon(T^2 - \frac{T^3}{\sqrt{T^2+R^2}}) 
( |HV\rangle _{AB} \langle HV| + |VH\rangle _{AB} \langle VH|)
\end{aligned}
\end{equation}

For $\varepsilon\rightarrow 0$ we 
obtain a polarization-entangled states which is mixed due to decoherence in a preferred polarization basis. The entanglement of such states is extremely resilient to the colored noise and could also violate  Bell's inequalities  \cite{Cabe05}. The Table I summarizes the theoretical values of the concurrence $C$ and the probability $P$ of success for each step of the protocol as function of the coupling strength $T$.




Let us now describe the experimental implemetation of the localization process. In Table II, we compare the theoretical predictions to experimental results for each step of the protocol.

\begin{figure}[h]
\centering
\includegraphics[width=0.26
\textwidth]{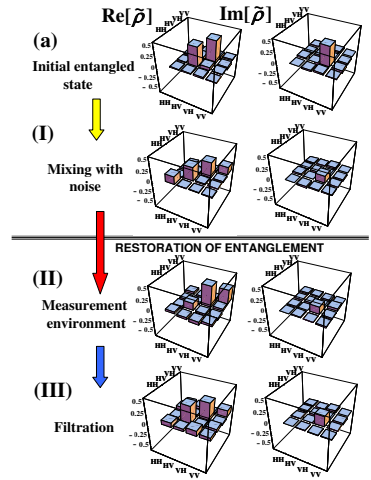}
\caption{Experimental density matrix for distinguishable photons: \textbf{a}%
) $\widetilde{\protect\rho }_{AB}^{in}$, \textbf{I}) $\widetilde{\protect%
\rho }_{AB}^{I}$, \textbf{II}) $\widetilde{\protect\rho }_{AB}^{II}$,
\textbf{III}) $\widetilde{\protect\rho }_{AB}^{III}.$ We reconstruct the two
qubit density matrix $\protect\rho _{AB}$ through the quantum state
tomography procedure \protect\cite{Jame01}. For each tomographic setting the
measurement lasts from 5 s ($\mathbf{a}$) to 30 minutes ($\mathbf{III}$),
the latter case corresponding to about $500$ triple coincidence. }
\end{figure}

\textbf{I)} After the mixing at the beam splitter with the completely distinguishable and unpolarized photon $E$, the singlet state $|\Psi ^{-}\rangle
_{AB}$ evolves into the density matrix ${\rho }_{AB}^{I}$ (see Table I). If there is just a single photon outgoing towards Bob's detection, the channel acts as
a depolarizing Pauli channel. As it can be observed in Fig.(3), for $T\leq \sqrt{2}-1$ the concurrence $C_{I}$, quantifying entanglement between photons in modes $k_A$ and $k_{B}$,
vanishes. To test the theory, we carried out an experiment by adopting a
beam splitter with $T\approx 0.4$. The experimental density matrix $%
\widetilde{\rho }_{AB}^{I}$ is shown in Fig.(4-\textbf{I}) and has a high
fidelity with the theoretical prediction (see Table II). The state obtained
experimentally exhibits $\widetilde{C}_{I}=0$, hence Alice and Bob do not share entanglement. 
\begin{table}[b!!]
\begin{center}
{\footnotesize
\begin{tabular}{||c|c||c|c||c||}
\hline\hline
$$ & $$ & $Theory$ & $Exp.$ & $Fidelity$\\ 
\hline\hline
$I$ & $C_{I}$ & $0$ & $0$ & $0.997\pm 0.006$\\ \hline
$II$ & $C_{II}$ & $0.32$ & $0.19\pm 0.02$ & $0.96\pm 0.03$\\
$$ & $P_{II}$ & $0.27$ & $0.26\pm 0.01$ & $$\\ \hline
$III$ & $C_{III}$ & $0.42$ & $0.28\pm 0.02$ & $0.89\pm 0.06$ \\ 
$$ & $P_{III}$ & $0.17$ & $0.11\pm 0.01$ & $$\\ \hline \hline
\end{tabular}
}
\end{center}
\caption{Theoretical and experimental data in the distinguishable surrounding photon regime for the three steps of the protocol \cite{Acc}.}
\end{table}
\textbf{II)}. The state evolves into an entangled one described by the
density matrix ${\rho }_{AB}^{II}$. Theoretically, the entanglement is
localized between Alice and Bob for all the values of $T\neq 0$ but the elements of the matrix are
fairly unbalanced, as shown by the experimental one $\widetilde{\rho }_{AB}^{II}$
reported in Fig.(4-\textbf{II}). \textbf{III)} If the transmittivity T, which characterizes the coupling, is known then the entanglement can be enhanced by filtration. The filters $F_{A}$ and $F_{B}$, acting on the
modes $k_{A}$ and $k_{B}$, ensures the lowering of the $|VV\rangle \langle VV|$ component, leading to a higher
concurrence, as can be seen in Fig.(3). The concurrence has a limit for asymptotic filtration ($%
\varepsilon \rightarrow {0}$) lower than unity (Fig.3-continuous line): maximal entanglement
cannot be approached, as a consequence of the coupling with incoherent photon.
Of course, a collective protocol can be still used \cite{Pan03},
since all entangled two-qubit state are distillable to a singlet one.
Applying the filtration with the attenuation parameters over the two modes equal to $A_{A}=0.33$ and $A_{B}=1$, we
obtain the state shown in Fig.(4-\textbf{III}). Such filtration brings to a
higher concurrence. The filtration also allows to produce 
a state violating Bell inequalities \cite{Verst02}, see Table I, although it was impossible just after the measurement. 

\begin{table}[b!!]
\begin{center}
{\footnotesize
\begin{tabular}{||c|c||c|c||c||}
\hline\hline
$$ & $$ & $Theory$ & $Exp.$ & $Fidelity$\\ 
\hline\hline
$I$ & $C_{I}^{\prime}$ & $0$ & $0$ & $0.86\pm 0.02$\\ \hline
$II$ & $C_{II}^{\prime}$ & $0.22$ & $0.15\pm 0.03$ & $0.96\pm 0.01$\\
$$ & $P_{II}^{\prime}$ & $0.20$ & $0.22\pm 0.01$ & $$\\ \hline
$III$ & $C_{III}^{\prime}$ & $0.47$ & $0.50\pm 0.01$ & $0.92\pm 0.04$ \\ 
$$ & $P_{III}^{\prime}$ & $0.09$ & $0.10\pm 0.01$ & $$\\ \hline \hline
\end{tabular}
}
\end{center}
\caption{Comparison between theory and experimental results in the partially indistinguishable regime \cite{Acc2} (Filtration (III) with parameters $A_{A}=0.12$, and $A_{B}=0.30$.}
\end{table}

\textbf{Partially indistinguishable surrounding photon:} 
As second scenario, we consider the state shared between Alice and Bob after
the coupling with the photon $E$, as a mixture arising from coupling with a
partially distinguishable noise photon. In such situation, the protocol exploits quantum interference
phenomena in the coupling allowing a higher localization of entanglement. The degree of indistinguishability
is parameterized by the probability $p$ that the fully depolarized
environmental photon is completely indistinguishable from signal. From the theory, the
entanglement between Alice and Bob disappears if $p>\frac{T^2+2T-1}{2T(1-T)}$. After the localization,
the concurrence after the measurement is $C_{II}^{\prime}=
\frac{T|T-p(1-T)|}{1-(2+p)T(1-T)}$ which is positive if $%
p\not= T/(1-T)$. After the application of the filtering, the concurrence can be
enhanced up to $C_{III}^{\prime}=\frac{|T-p(1-T)|}{\sqrt{%
1-2(1+p)T(1-T)}}$ which is higher than $C_{III}$.  Hence it is advantageous to induce
indistinguishability between the noise and the signal. In the theoretical limit of
asymptotic filtration and completely indistinguishable photons ($p=1$),
the concurrence reaches unity except for $T=0,1/2$ \cite{tele}. In order to ensure an indistinguishability between the photons $B$ and $E$ the mutual delay $\Delta t$ has been set equal to $0$, corresponding  to the maximal
overlap of the photon pulses injected into $BS$. Furthermore
 on modes $k_{B}$ and $k_{E}$ we have inserted narrow-band interference filters with $\Delta\lambda=1.5nm$ (Fig.2). We have
estimated the degree of indistinguishability between the photon belonging to
mode $k_{B}$ and the one associated to $k_{E}$ as $p=(0.85\pm 0.05)$ by
realizing an Hong-Ou-Mandel interferometer \cite{Hong87} adopting a beam splitter with $%
T=0.5$. To carry out the experiment we adopt a $BS$ with $T=0.3$ and the
optical delay has been set in the position $\Delta t=0$ to reach maximal interference at the beam splitter. 
According to the data reported in Table III, for partially
indistinguishablity from the environmental photon, the interference phenomena
taking place in the coupling process allows to localize a higher amount of
initial entanglement.

In summary, we have investigated the role of quantum coherence in the entanglement localization protocol, which is able to restore the entanglement although it has been redirected by a coupling to the surrounding noisy and incoherent system. We have reported the experimental proof-of-principle demonstration of the protocol localizing polarization entanglement between two photons after the coupling with distinguishable (incoherent) and unpolarized photon. It has been already justified that this localization protocol works also for a multiple linear couplings with many incoherent photons,
representing a more complex surrounding system \cite{Naga08}. 
A direct application of this entanglement localization can be envisaged to improve 
generation of entanglement, in-line quantum communication and in computation.

R.F. and M.G. have been supported by MSM 6198959213
and LC 06007 of Czech Ministry of Education, Grant
202/08/0224 of GACR and AvH Foundation.


\begin{thebibliography}{99}
\bibitem{Gen}  All the present results can be directly extended for general
passive coupling between the two modes.
\bibitem{Acc}  In our experiment, the environmental field state led to
non-negligeable contributions of 2 photons. These ones have been estimated
from the detector count rates and have been subtracted from the experimental
data. Without substraction we found $C_{II}=0.15$ and $C_{III}=0.20$.
\bibitem{tele} For a fully coherent coupling scheme the signal and environmental noise are indistinguishable. Hence  the entanglement is redistributed between the three particles and it is localized back through measurement and filtration. However, at variance with the teleportation protocol, the environmental photon is in a fully mixed state.
\end{thebibliography}
\end{document}